\documentclass[aps,prb,twocolumn,floats]{revtex4}
\usepackage{epsfig}
\usepackage{amsmath}
\usepackage{graphicx}
\usepackage{color}

\begin{document}
\title{Chiral anomaly induced magnetoconductances in an irradiated Type-I Weyl Semimetal}
\author{Rounak Sen, Satyaki Kar$^*$}
\affiliation{AKPC Mahavidyalaya, Bengai, West Bengal -712611, India}
\begin{abstract}
  Magneto conductivities in Weyl semimetals (WSM) in presence of small fields are studied using quasi-classical Boltzmann transport equations (BTE). Following such formalism here we consider irradiation via circularly polarized light on a two-node time reversal breaking WSM already under a dc/static electric field and study the magneto-transport properties due to the presence of chiral anomaly. Chiral anomaly affects both longitudinal magnetoconductivity as well as planar Hall conductivity. {As our field set-up causes continuous time variation in the relative orientation between the fields, one naturally expects interesting magneto-transport behavior for different field strengths and tilting.} The type-I tilting that we study here displays both positive and negative magnetoconductances depending on the field strengths and time. Furthermore, we find that a direct temporal tuning of the irradiated field strengths can {lead to fluctuating} magneto-transport behavior which can be easily improvised and checked in the laboratories.
\end{abstract}
\maketitle                              
\section{Introduction}
Recent trends in condensed matter research revolves a lot around topology\cite{neto,hasan,zhang} and at the heart of this topological condensed matter field lies the recently developed Weyl semimetal (WSM)\cite{aswin,rao,felser,hasan2,zou,shen,jayanavar} - the materials that feature Weyl fermions in momentum space in the form of Berry curvatures with monopoles\cite{suyang}. These materials show high electron mobility and magnetoresistances\cite{sekhar}. But similarly interesting is its response to electromagnetic fields. With electric and magnetic field in three dimension, such materials display the exotic phenomena called Adler-Bell-Jackiw chiral anomaly\cite{aswin,rao,felser,hasan2,zou,shen,jayanavar,zhang2,reis,jia}.
It refers to the non-conservation of chiral charges in individual Weyl nodes of the Weyl semimetals when non-orthogonal electric and magnetic fields are applied in such systems. This causes a positive longitudinal magnetoconductance (LMC) that goes as $B^2$ for small magnetic field $B$\cite{reis,girish96,prl119}. It also results in planar Hall effect (PHE) if there are in-planar non-parallel electric and magnetic fields\cite{girish96,prl119}. Even with a magnetic field alone, such systems can show chiral magnetic effect\cite{aswin,son,kim} where simple derivation using Boltzmann transport equation\cite{girish96,prl119,son,kim,nagaosa,zyuzin} (BTE) shows a chiral current proportional to the chemical potential difference at the opposite-chirality Weyl nodes. 

It has been found that it is due to the Berry curvature that the semiclassical equations of motion gets modified incorporating additional terms amounting to anomalous Hall effect, chiral magnetic effect and chiral anomaly. Considering a simple time reversal breaking (TRB) two-node WSM and applying BTE around each of those nodes, one can get the equation of continuity involving number density for charge carriers $N^\pm$ and current $j^\pm$ as\cite{son,kim}
\begin{eqnarray}
\partial N^\pm/\partial t+\nabla_r.j^\pm={\rm c}^\pm\frac{e^2}{4\pi^2}E.B
\label{eq1}
\end{eqnarray}
where $N^\pm=\int_{-\infty}^\infty d\epsilon\rho^\pm(\epsilon)f^\pm(\epsilon,r,t)$ and $j^\pm=\int \frac{d^3p}{(2\pi)^3}[v+eE\times\Omega^\pm+\frac{e}{c}\Omega^\pm.vB]f^\pm(p,r,t)$. Here $\epsilon$ denotes dispersion energy and $E$ and $B$ denote the electric and magnetic fields respectively. Also $\rho^\pm(\epsilon)=\int\frac{d^3p}{(2\pi)^3}(1+\frac{e}{c}B.\Omega^\pm)\delta(\epsilon_p-\epsilon)$ denote the density of states, $f^\pm$ the Fermi distribution function, $\Omega$ the Berry curvature and ${\rm c}^\pm=\frac{1}{2\pi}\int dS.\Omega^\pm$ denote the Chern numbers at the $\pm$ Weyl nodes. This leads to
\begin{eqnarray}
\partial (N^+-N^-)/\partial t+\nabla_r.(j^+-j^-)=\frac{e^2}{2\pi^2}E.B
\label{eq2}
\end{eqnarray}
which is essentially the chiral current nonconservation due to chiral anomaly.
We consider chemical potential to be away from Weyl nodes, $i.e.,$ $\mu>>k_BT,~\hbar\omega_c$ ($\omega_c=eB/m$). The scattering is considered to be elastic and the collision integral $I_{coll}$ depends on elastic relaxation times $\tau_{intr}$ and $\tau_{inter}$ (for intra and inter node scattering respectively).
For $\tau_{intr}<<\tau_{inter}$, the electron distribution function around WP's depend on $\epsilon$ alone\cite{son} and considering homogeneity of the system a relaxation time approximation is adopted in the scattering integral involving $\tau_{inter}~(=\tau$, say) alone\cite{son}.

In presence of small/weak fields, Landau quantization gets wiped out\cite{girish96,prl119} and the Boltzmann formalism remains sufficient to describe the magneto-transport. This results in negative magnetoresistances in the WSMs due to the chiral anomaly which has been widely reported from experiments as well\cite{huang,li}. The longitudinal magnetoresistance LMR = $\frac{\rho_{xx}(B)-\rho_{xx}(0)}{\rho_{xx}(0)}\times100\%$ becomes negative when resistivity $\rho_{xx}(B)$ (or conductivity $\sigma^{xx}$) decreases (increases) with $B$, where $x$ denotes the longitudinal direction ($i.e.,$ along $E$).
The anomaly related longitudinal magnetoconductance shows positive $B^2$ dependence in a two node WSM when $B||E$ and normal to the Weyl node directions\cite{girish96}. The corresponding contribution to the anomaly related LMR remains negative in presence of Berry curvature $\Omega_p$ and does not cancel out even when opposite chirality nodes are enclosed in a single Fermi surface\cite{nagaosa}. However, LMC shows a linear B dependence if $E,~B$ are directed along the nodes or tilt direction\cite{girish96,prl119}. It thus needs thorough investigation on the exact behavior of LMC in terms of B at an arbitrary longitudinal orientation.

 In light irradiated systems, the fields can change orientation as well as strengths with time. And it's important to analyze the magnetoconductivities in those situations. In this paper, we consider a periodic variation of fields and study their effect on the magneto-transport.
Notice that the LMC in an untilted WSM with ${\hat x}$ being the longitudinal direction can be expressed\cite{prl119} as
\begin{eqnarray}
  \sigma^{xx}=\sigma+\Delta\sigma cos^2\theta_{eb}\\\nonumber
  \sigma^{yx}_{phe}=\Delta\sigma sin\theta_{eb} cos\theta_{eb}
  \end{eqnarray}
where $\theta_{eb}$ is the angle between $E$ and $B$ field and $\Delta\sigma\propto B^2$. Thus if we have a $B\sim cos(\omega t)$ dependence, the LMC shows temporal fluctuations but it never becomes negative. But for $\Delta\sigma\propto B$, one can certainly have a negative LMC. One can also have oppositely directed PHE depending on the field orientations. With tilting such simple relations get modified and a negative LMC can be observed at small fields depending on the values of tilt parameter and Fermi energy. There the magneto-transport also shows nontrivial features if $\theta_{eb}$ possess time dependence in addition.
This paper deals with various of such possibilities and investigate closely the features of chiral anomaly induced conductivities in WSM systems in presence of time periodic fields.

In section II, we formulate the continuum Hamiltonian for a two-node WSM and the general expressions for LMC and PHE. In section III, {we consider irradiation via circularly polarized laser onto such systems} in addition to a dc electric field and we discuss the results on magnetoconductivities in detail. Finally in section IV, we summarize our findings and discuss their utilities and drawbacks and possible future directions.

\section{Formulation}
A typical tilted WSM Hamiltonian\cite{zyuzin} can be written as 
\begin{eqnarray}
H=\hbar sv(k_x\sigma_x+k_y\sigma_y+k_z^{s}\sigma_z)+\hbar sCvk_z^{s}
\end{eqnarray}
where $k_z^{s}=k_z-sQ$ in a two-node WSM with Weyl nodes at $(0,0,sQ)$ with $s=\pm1$ denoting two Weyl points (WP) and $C$ is the tilt parameter. We only consider a type I WSM for which $|C|<1$. Once we introduce electric field $E$ and small magnetic field $B$ in this system, chiral flow (and hence anomaly) develops. We keep our $E$ and $B$ fields in the $x-z$ plane to study LMC and PHE.
First of all we need to probe whether low $B$ LMC shows a linear $B$ dependence due to tilt in the system.
Using relaxation time approximation one can write $\sigma_{L}$ ($L$ stands for longitudinal and indicates direction parallel to current, which will be along the nodal axes) as\cite{prl119}
\begin{eqnarray}
\sigma_{L}={e^2\tau}\sum_s\int \frac{d^3k}{(2\pi)^3}\{D[v_{L}+\frac{eB~cos\theta_{eb}}{\hbar}{\bf v.\Omega}]^2(-\frac{\partial f_{eq}}{\partial\epsilon})
\label{sigL}
\end{eqnarray}
Here ${\bf \Omega}=-s\frac{\bf k}{k^3}$ is the Berry curvature, $D=[1+\frac{e}{\hbar}{\bf B.\Omega}]^{-1}$ is the phase space factor, $f_{eq}$ is the equilibrium distribution function and $v_{L}=\partial\epsilon/\partial k_{L}$ with dispersion $\epsilon^\pm=\pm\hbar  v\sqrt{k_x^2+k_y^2+(k_z^{s})^2}+\hbar vsCk_z^{s}$ for conduction (+) and valence (-) bands respectively. We focus only on the conduction band and call $\epsilon=\epsilon^+$.
We see the $v_L$ does not depend on $C$ if the direction of $E$ is normal to the nodal axes or ${\hat z}$ direction. Additionally if $B$ has nonzero component along $E$ ($i.e., \theta_{eb}\ne\pi/2$), $\sigma_{L}$ grows quadratically with $B$. But if  ${\bf E}=E\hat z$ and $\theta_{eb}\ne\pi/2$, it is only at nonzero tilting, the linear in $B$ term survives and dominates for small $B$ values.

Now let's take a look at the planar Hall conductivity $\sigma^{phe}$ (with both $E$ and $B$ in the $xz$ plane). This is given as
\begin{align}
\sigma^{phe}&=e^2\tau\sum_s\int \frac{d^3k}{(2\pi)^3}D[v_{L}v_\perp+\frac{eB}{\hbar}(v_Lsin\theta_{eb}+v_\perp cos\theta_{eb})\nonumber\\
    &*{\bf v.\Omega}+\frac{e^2B^2}{\hbar^2}sin\theta_{eb} cos\theta_{eb} ({\bf v.\Omega})^2](-\frac{\partial f_{eq}}{\partial\epsilon}).
\label{sigxy}
\end{align}
It implies a $B^2$ dependence unless $\theta_{eb}=0,\pi/2$ when a linear $B$ dependence is observed in $\sigma^{phe}$.

\begin{figure}
  \includegraphics[width=.6\linewidth]{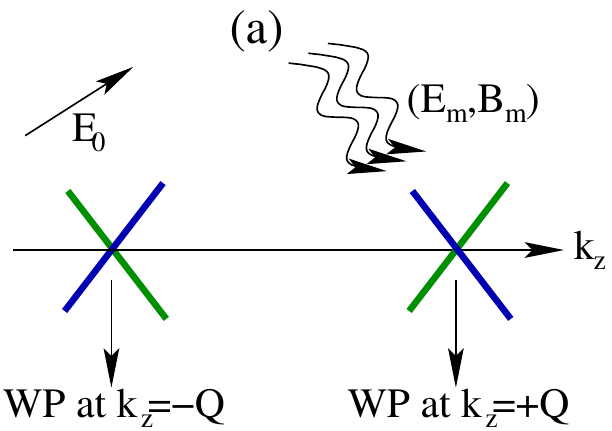}
  \vskip .25 in
\includegraphics[width=.45\linewidth]{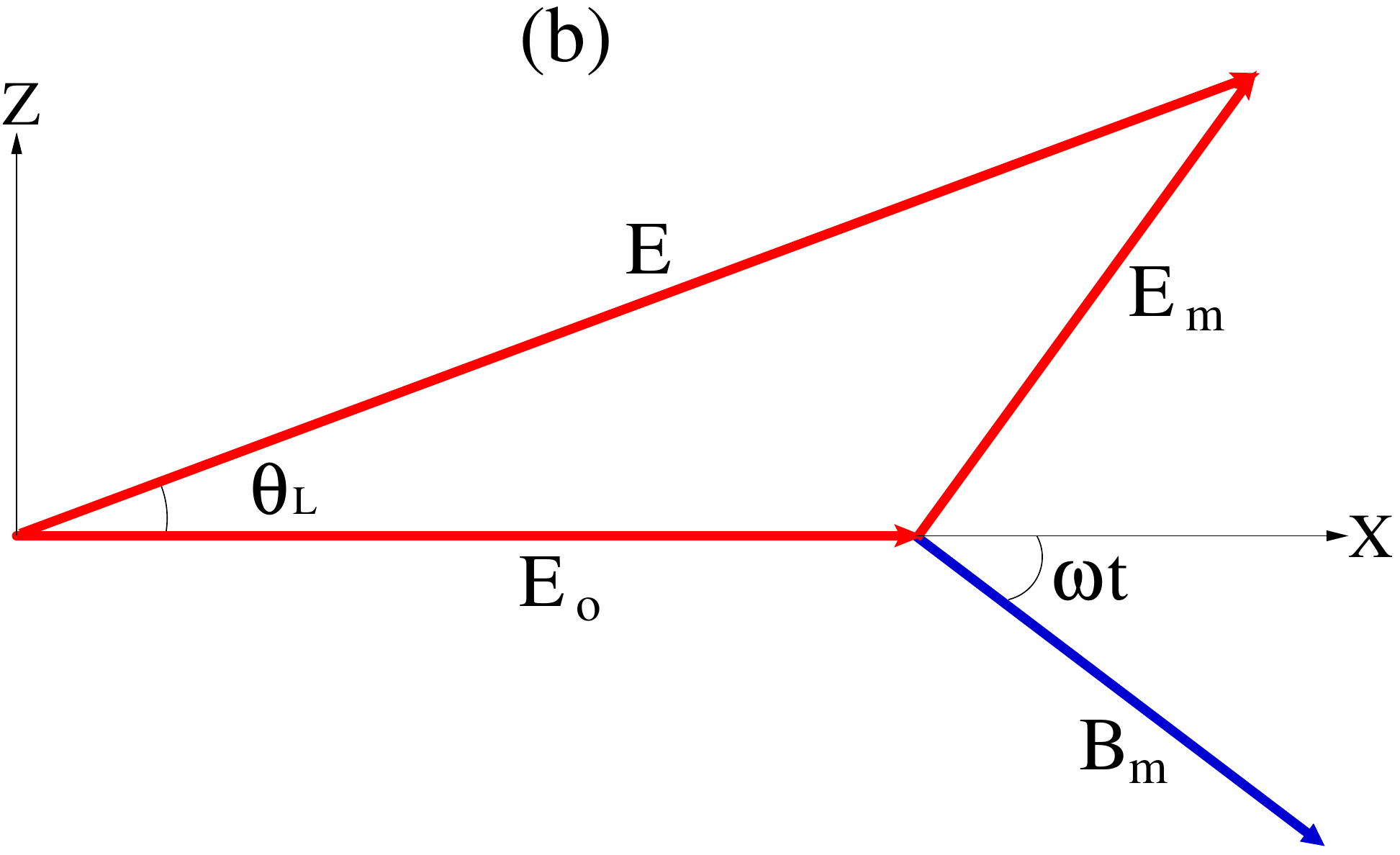}
  \includegraphics[width=0.45\linewidth,height= .9 in]{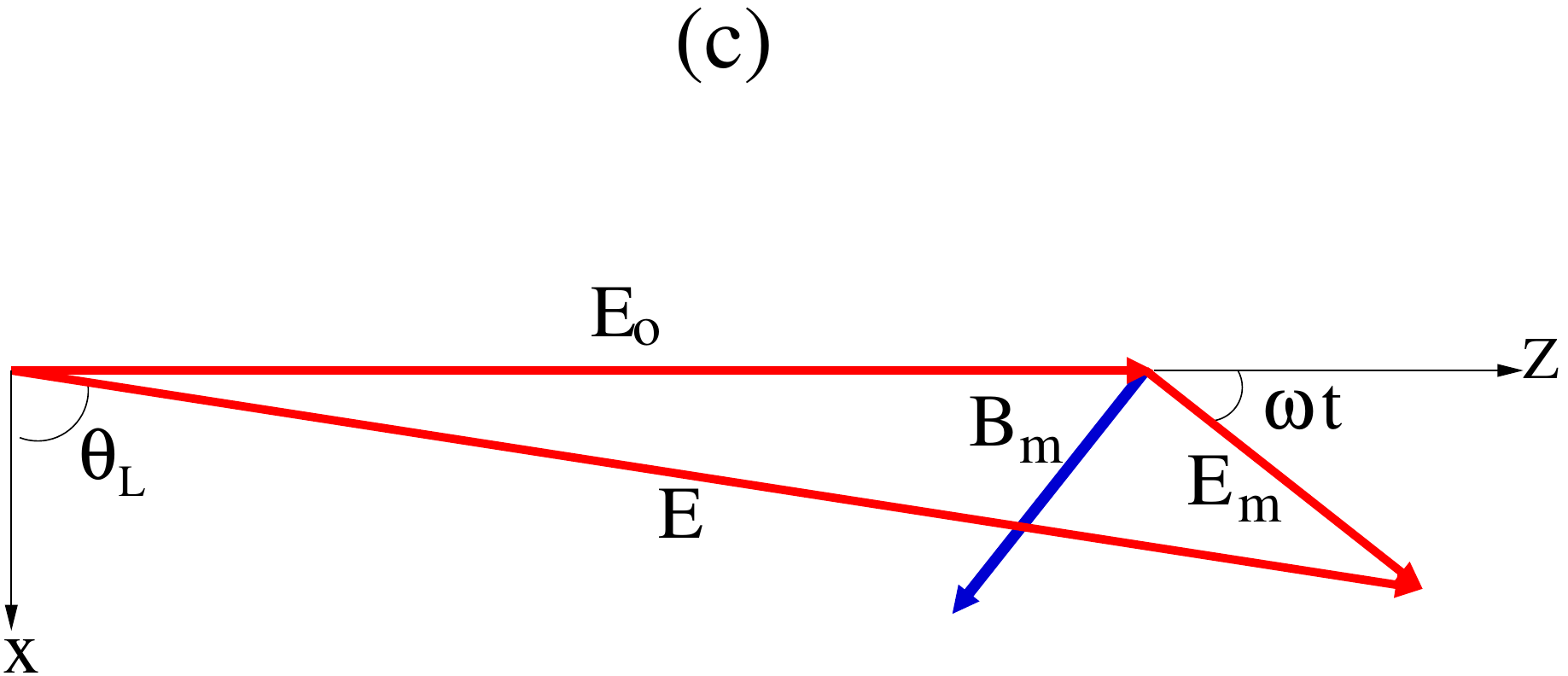}
\caption{(a): Pair of Weyl points in presence of a dc/static field $\hat E_0$ and ac fields $(E_m,B_m)$. (b),(c): Combination of static and electromagnetic fields with $\hat E_0$ along (b) $\hat x$ and (c) $\hat z$.} 
\label{cartoon}
\end{figure}


\section{Irradiation via EM wave with circular polarization}
We find that a combination of dc and ac electric field can produce interesting time dependent magnetoconductivity phenomena in simple two-node WSM systems.
The longitudinal direction being the electric field direction, it changes periodically and the chiral anomaly effect causes conductivities to change accordingly with tilting being another major tuning parameter.

  In our set up, we first consider a large static electric field $E=\hat E_0=E_0\hat x$. Then the such system is exposed to irradiation via a circularly polarized light with an electric field $E=E_m(sin \omega t, 0,cos\omega t)$ and a magnetic field $B_m(cos\omega t, 0, -sin\omega t)$ which are interrelated via Maxwell's equations as $E_m=cB_m$ (see Fig.\ref{cartoon}). When $E_m/E_0<<1$, we can call $\hat x$ direction to be the longitudinal direction and accordingly the temporal variation of $\sigma_L$ and $\sigma_{phe}$ can be shown to be a periodic function. But in general we call $\theta_L=tan^{-1}\frac{E_mcos(\omega t)}{E_0+E_msin(\omega t)}$ to denote the longitudinal direction ($i.e.,$ the direction of E from $\hat x$ axis) and study the periodic variation of LMC with time, particularly for $C\ne 0$.  We take $E_0$ to be 10 gigavolt/m and {usually vary $B_m$ from 0-5 T. Such alternating fields can be provided by irradiation via intense lasers (with intensity upto $\sim 10^{19}~W/cm^2$)\cite{laser} and are often used for comparison with Floquet theory based calculations\cite{banasri,flq1,flq2}.} In this paper, the fields $B_m$ and conductivities $\sigma_L$ and $\sigma^{phe}$ are obtained in units of Tesla and Siemens/meter respectively.
With tilting, the dispersion becomes $\epsilon=\hbar vk(1+sCcos\theta)$. Thus the velocity components become $v_k=v(1+sCcos\theta)$ and $v_\theta=-vsCsin\theta$ with ${\bf v}=v_k{\hat k}+v_\theta{\hat\theta}$ where ${\hat k}=sin\theta cos\phi{\hat x}+sin\theta sin\phi{\hat y}+cos\theta{\hat z}$ and ${\hat \theta}=cos\theta cos\phi{\hat x}+cos\theta sin\phi{\hat y}-sin\theta{\hat z}$ in a spherical coordinate system.
The Longitudinal direction can be given as ${\bf\hat L}=cos\theta_L\hat x+sin\theta_L\hat z$ (and the transverse direction ${\bf\hat \perp}= -sin\theta_L\hat x+cos\theta_L\hat z$). This makes $v_L={\bf v.\hat L}=v[cos\theta_Lsin\theta cos\phi+sin\theta_L(cos\theta+sC)]$ and similarly $v_\perp=v[-sin\theta_Lsin\theta cos\phi+cos\theta_L(cos\theta+sC)]$. The magnetic field direction is given by $\theta_B=-\omega t$ and $\theta_{eb}=\theta_L-\theta_B=\theta_L+\omega t$.
With $C$, the $x$ and $z$ components of velocity becomes $v_x={\bf v.{\hat x}}=vsin(\theta)cos(\phi)$ and similarly $v_z=v(cos\theta+sC)$. In addition, we get ${\bf v.\Omega}=-(1+sCcos\theta)sv/k^2$.

Now these are the expressions when the static field ${\hat E_0}=E_0\hat x$. We also consider the case where the static field orients along nodal direction, $i.e.,~E=E_0{\hat z}$. This alters the longitudinal orientation causing $\theta_L=tan^{-1}\frac{E_0+E_mcos(\omega t)}{E_msin(\omega t)}$. The other expressions for conductivities, however, remain the same functions of $\theta_L$.

\subsection{Longitudinal Magnetoconductance}

\begin{figure}
\includegraphics[width=.48\linewidth,height=1.4 in]{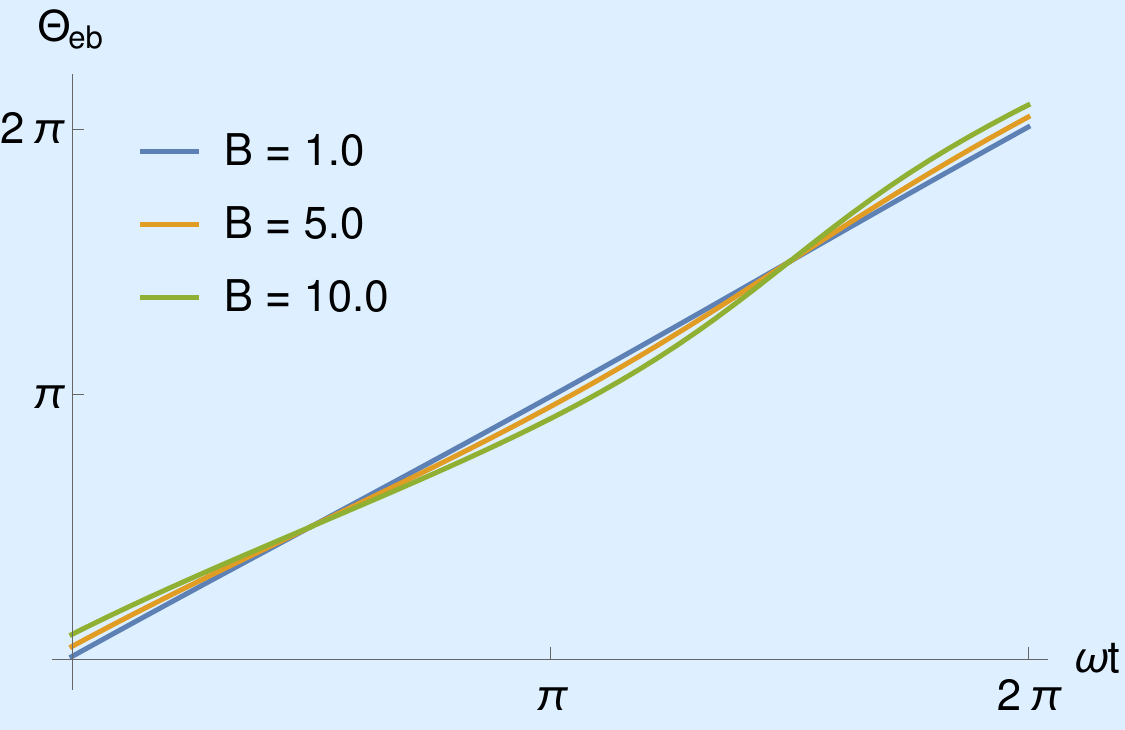}
  \includegraphics[width=0.48\linewidth,height=1.4 in]{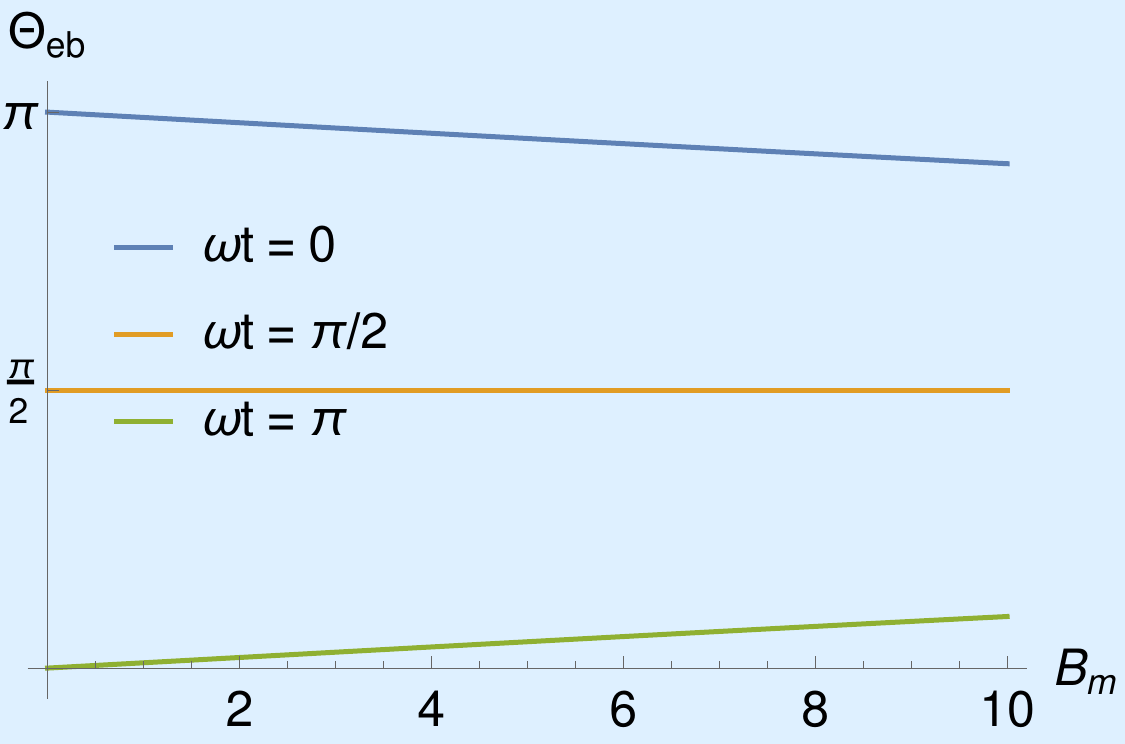}
\caption{(Color online) Variation of $\theta_{eb}$ with (left) $\omega t$ and (right) $B$.} 
\label{theta-eb}
\end{figure}

Let us first consider the effect on $\sigma_L$. First notice that $\theta_{eb}$ varies both with $t$ and $B_m$ (see Fig.\ref{theta-eb}). {Note that the large field $B_m= 10~T$ is shown here only for demonstration purpose and not used henceforth in the paper.} $\theta_{eb}$ being time dependent, the response on conductivities often become nontrivial.
\begin{figure}
\includegraphics[width=.69\linewidth]{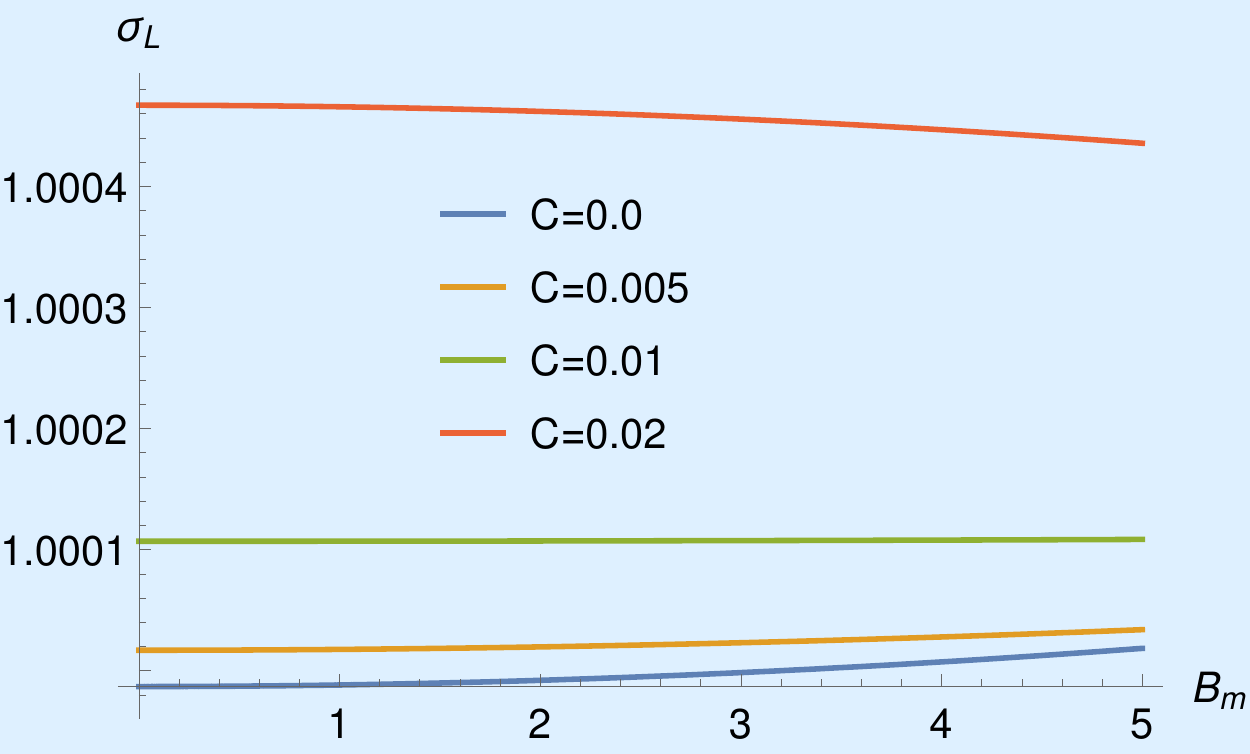}
  \includegraphics[width=0.69\linewidth]{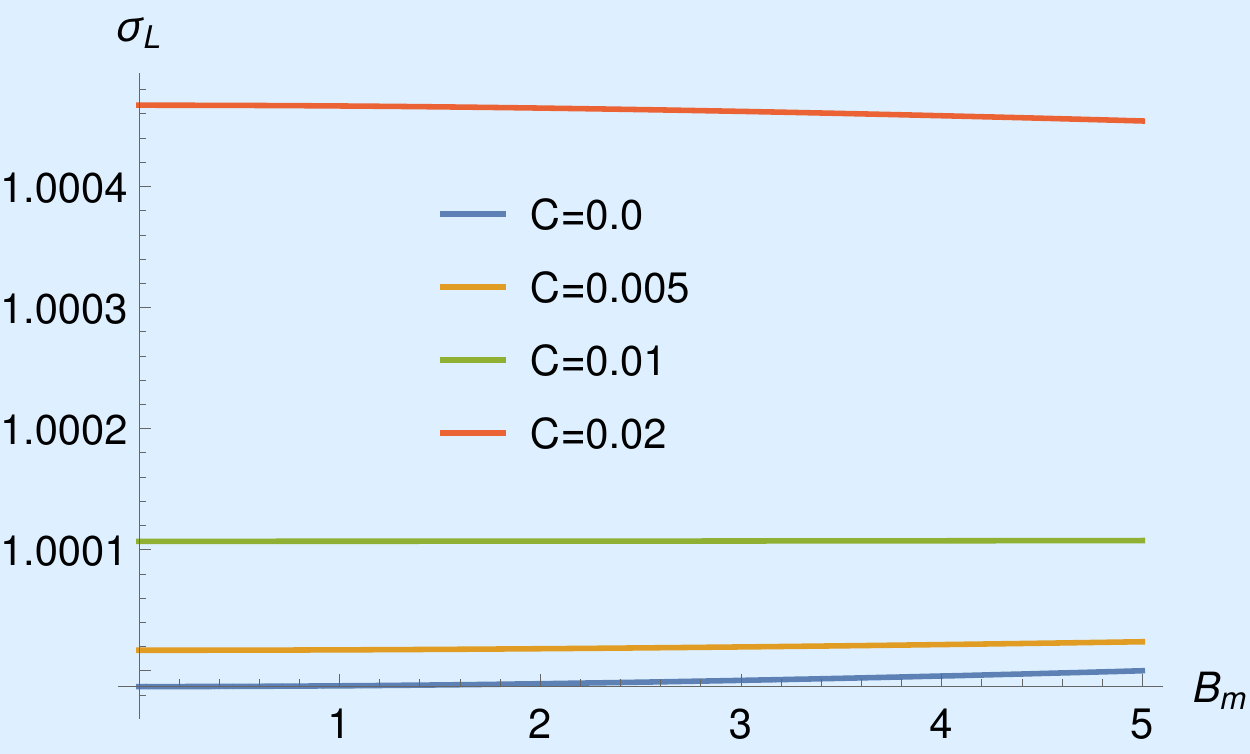}
\caption{(Color online) Variation of $\sigma_L$ with $B$ for $\omega t=0$ (top) and $\omega t=\pi/4$ (bottom) for different tilting parameters ($\sigma_L$'s normalized by their values at $B_m=0$ and $C=0.0$).} 
\label{LMCB}
\end{figure}

\begin{figure}
\includegraphics[width=.8\linewidth]{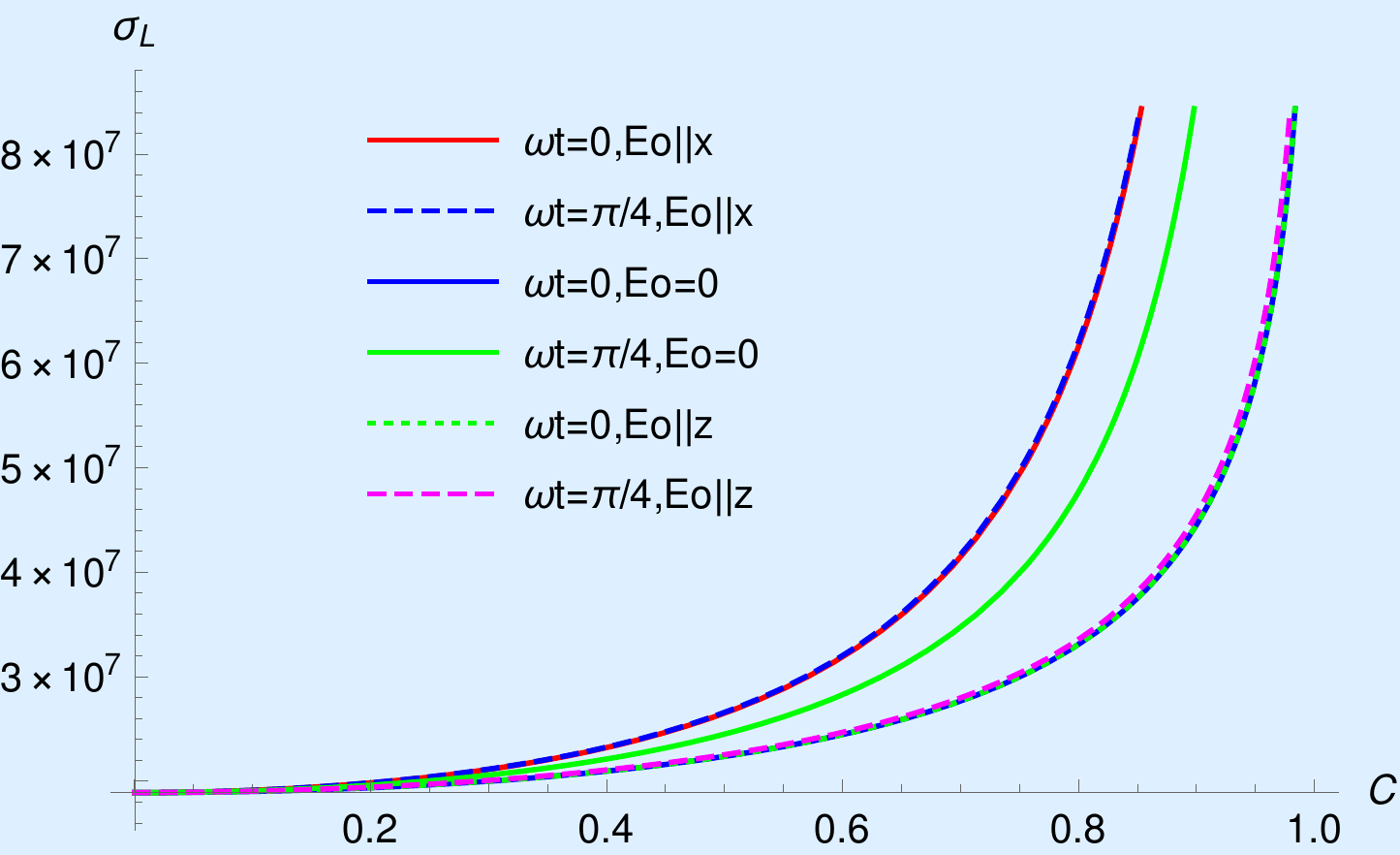}
 \caption{(Color online) Variation of $\sigma_L$ with $C$ for $\omega t=0$ and $\pi/4$ at $B_m= 5T$ with $E_0$ along $\hat x$, along $\hat z$ as well as with $E_0=0$. $\sigma_L$'s are in units of Siemens/meter.} 
\label{Call}
\end{figure}
\begin{figure}
\includegraphics[width=.48\linewidth,height=1.2 in]{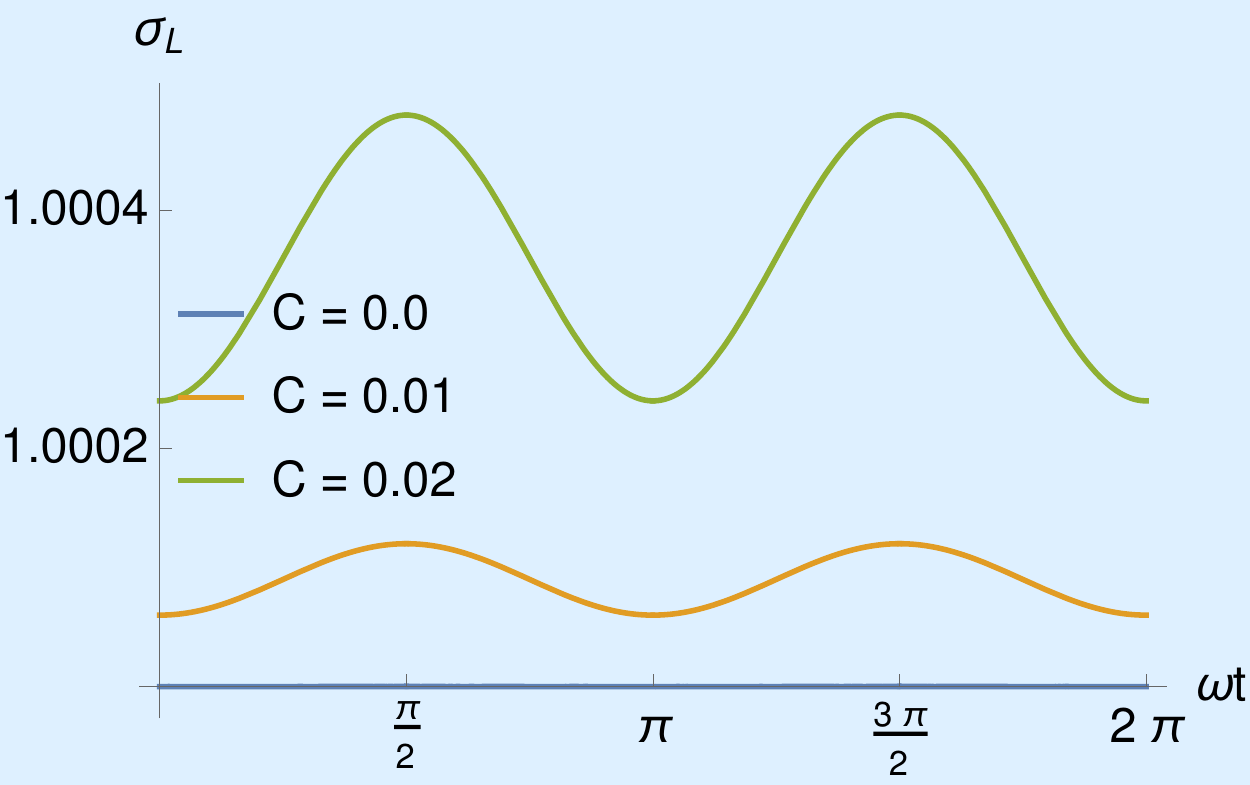}
  \includegraphics[width=0.48\linewidth,height=1.2 in]{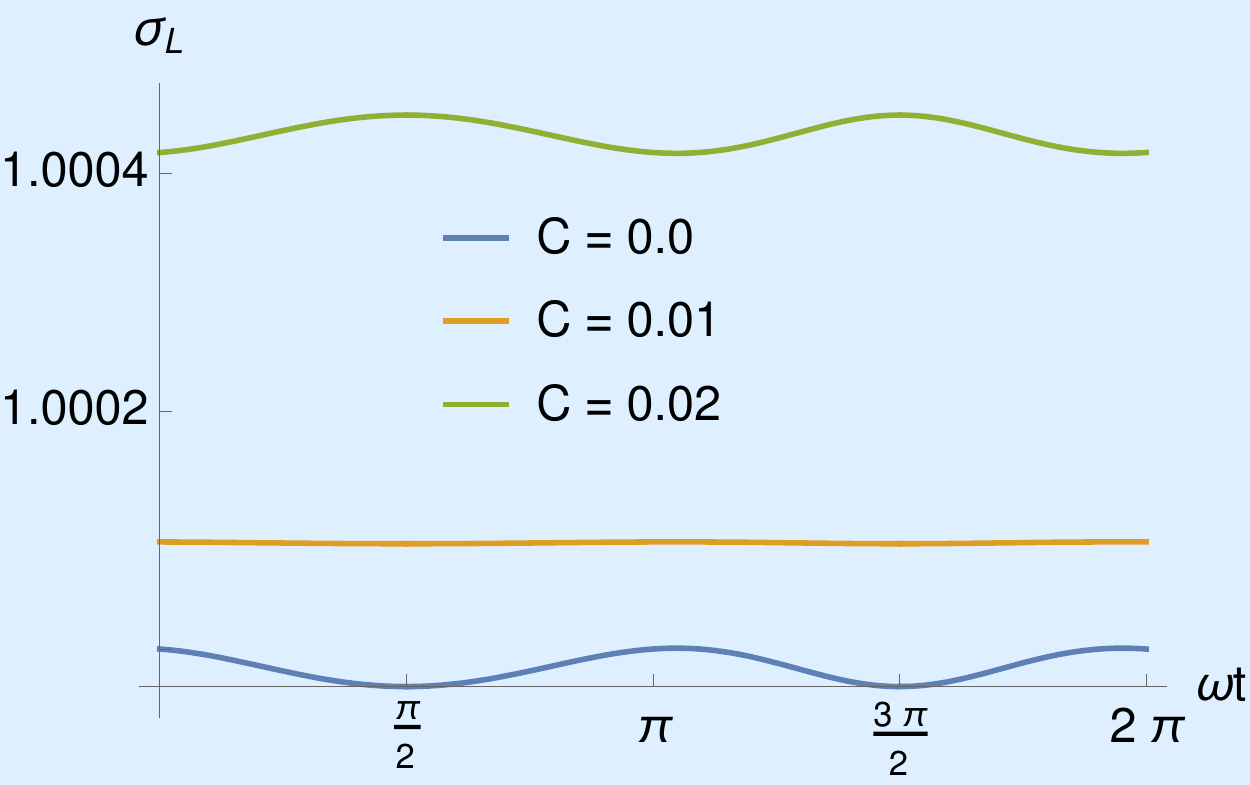}
\caption{(Color online) Variation of $\sigma_L$ with $\omega t$ for $E_0=0$ and $E_0\ne0$ at $B_m= 5 T$ for different tilting parameters ($\sigma_L$'s normalized by their values at $t=0$ and $C=0$).} 
\label{LMCt}
\end{figure}

In the following we show and describe the variations of $\sigma_L$ with $B_m$ (Fig.\ref{LMCB}), $C$  (Fig.\ref{Call}) and $\omega t$  (Fig.\ref{LMCt}). The results depend on the Fermi energy $\mu$. With $\mu$=1 eV, and for the large $E_0$ considered, $\sigma_L$ increases with $B$ at $C=0$ but as tilting is turned on and increased, {we see that even for very small nonzero tilting, $\sigma_L$ reverses the trend and start decreasing with $B$} implying a positive magnetoresistance, though all these changes are marginal compared to the magnitude of $\sigma_L$ (see Fig.\ref{LMCB}). {Notice that there is $C$ dependence in $\sigma_L$ even when $B_m=0$ (and thus $\theta_L=0$). This is because the spectrum depends on $C$ and so does the integral in Eq.\ref{sigL}.} We find that $\sigma_L$ increases with tilting with a higher rate as one reaches $C\rightarrow1$ (see Fig.\ref{Call}). Also for $C\ne0$, its dependence on $\omega t$ diminishes more and more with large $E_0/E_m$ considered (see Fig.\ref{LMCt}).

\begin{figure}
\includegraphics[width=.8\linewidth]{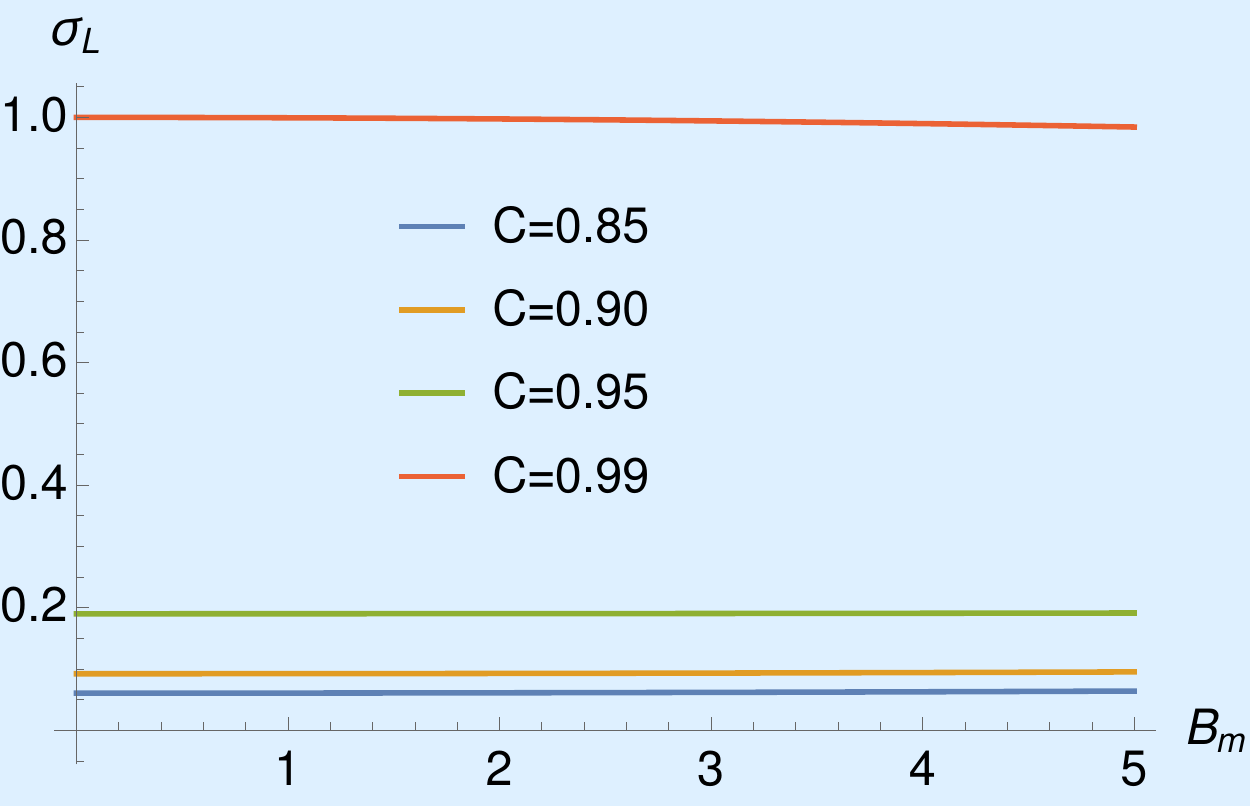}
\caption{(Color online) Variation of $\sigma_L$ with $B$ for $\mu=0.1$ eV at $\omega t=0$  for different tilting parameters ($\sigma_L$'s normalized by their values at $B_m=0$ and $C=0.99$).} 
\label{LMC.1}
\end{figure}
\begin{figure}
\includegraphics[width=.48\linewidth,height=1.2 in]{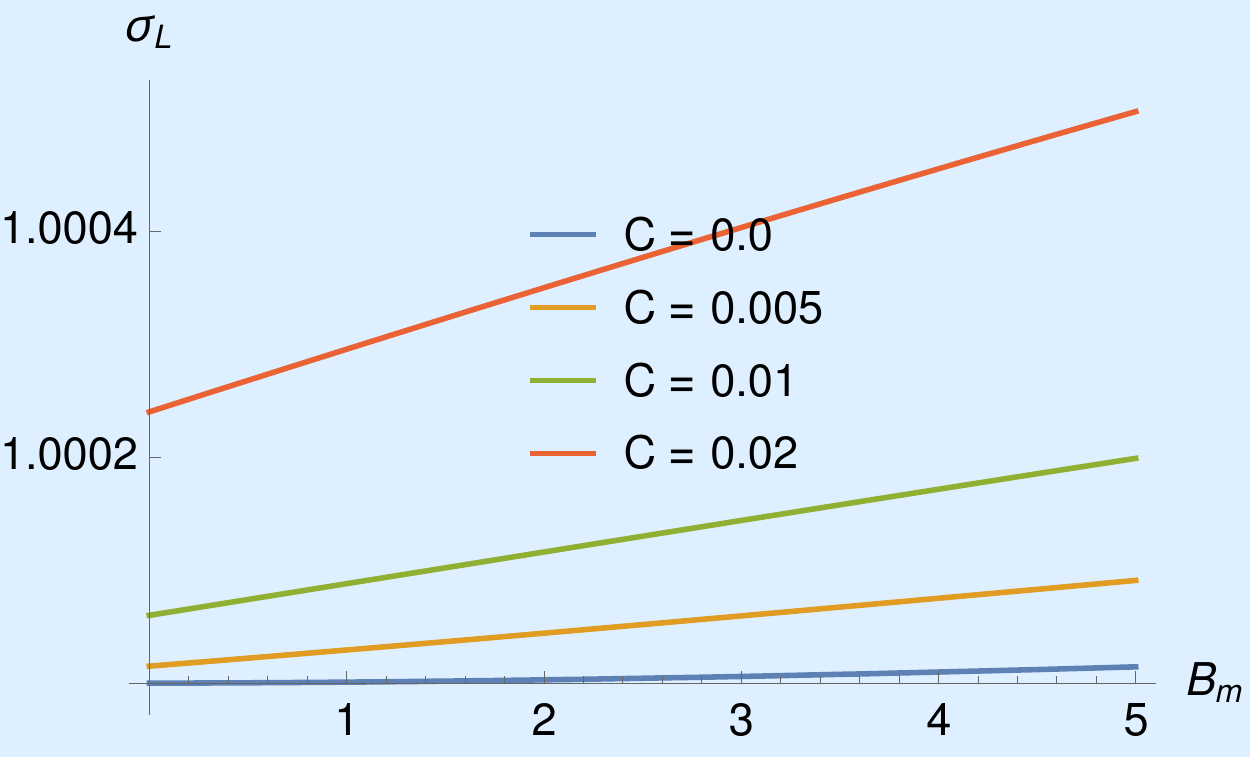}
\includegraphics[width=.48\linewidth,height=1.2 in]{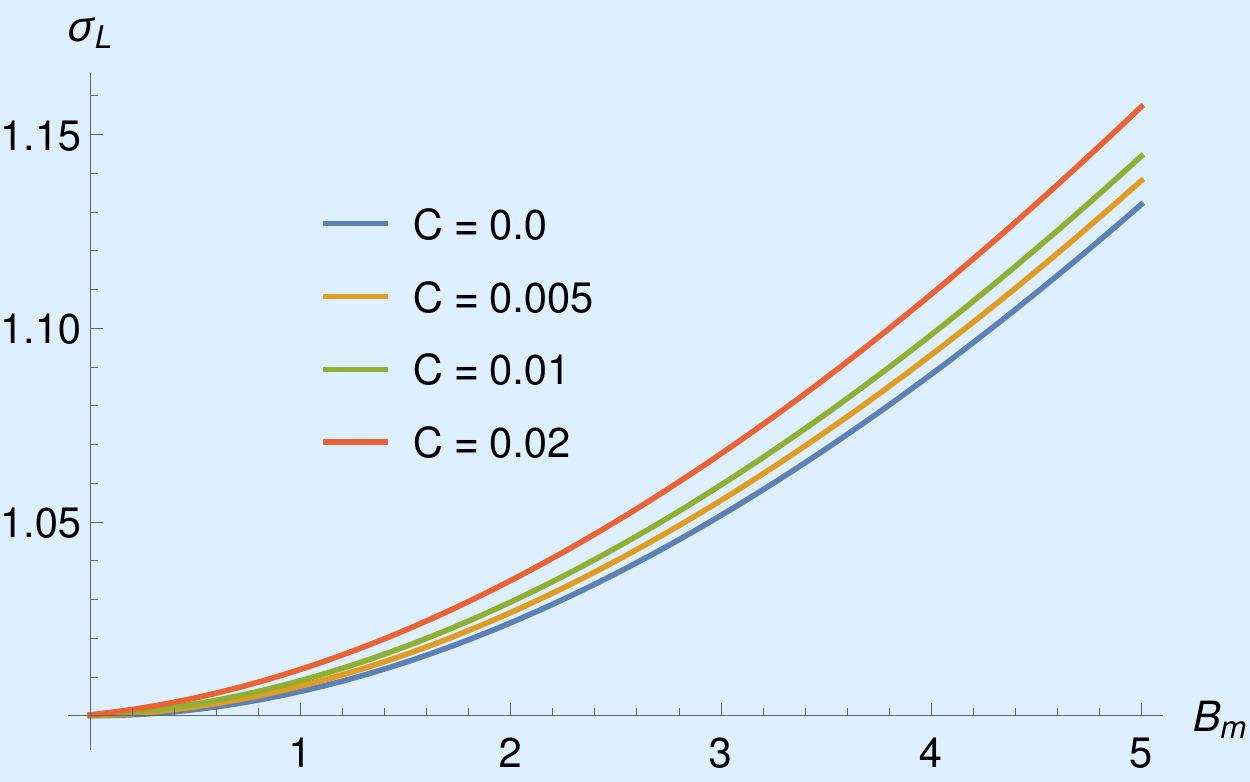}
\caption{(Color online) Variation of $\sigma_L$ with $B$ for $\mu=1$ eV (left) and $\mu=0.1$ eV (right) at $\omega t=\pi/4$ and with $E_0$ along $\hat z$ direction for different tilting parameters ($\sigma_L$'s normalized by their values at $B_m=0$ and $C=0$).} 
\label{EozLMC2}
\end{figure}

A follow up with dispersion results from WSM materials like $TaAs$, $TaP$, $NbAs$ or $NbP$\cite{taas,tap,felser} reveal that 1 eV is large enough for a linear $\epsilon-k$ relation. So we check our results for a smaller $\mu=0.1$ eV as well. And our results shows similar behavior though the magnetoconductance become negative only for $C\sim1$ (see Fig.\ref{LMC.1}).

Next we consider $\hat E_0$ to be along the $\hat z$ direction. There the variation of $\sigma_L$ with $B_m$ does not show negative magnetoconductance anymore. We find linear increase in $\sigma_L$ with $B_m$ for small fields at $\mu=1$ eV. But it actually depends on the relative strength of linear and quadratic terms of Eq.\ref{sigL}. For $\mu=0.1$ eV, we see the the quadratic term dominates over the linear term and the linear variation of $\sigma_L$ is not discernible within the window of 1T - 5T for $B_m$ that is shown in Fig.\ref{EozLMC2}. We also notice that $\sigma_L$ is much larger for large tilting (but still of type-I genre) if $E_0$ points along $\hat x$ instead of $\hat z$ (see Fig.\ref{Call}).

\subsection{Planar Hall Conductance}
Let us now find out how the above mentioned field variations affect the planar Hall conductance in WSM systems.
Notice that in Eq.\ref{sigxy}, the term linear in $B$ comes out to be $\frac{e^2\tau}{\pi^2}\frac{eBCv}{\hbar^2}cos[\theta_{eb}-\theta_L]$ because we consider both the Weyl nodes in obtaining the conductivity and thus vanishes for $C=0$. Similarly the $B$ independent term also vanishes for $C=0$. So in general, we can write $\sigma^{phe}=\sigma_0+aB+bB^2$, $\sigma_0$, $a$ and $b$ being constants. For $C=0$ we obtain $\sigma_0=a=0$ and $\sigma^{phe}$ becomes proportional to $B^2$.

\begin{figure}
  \vskip .1 in
\includegraphics[width=.39\linewidth,height= 1.2 in]{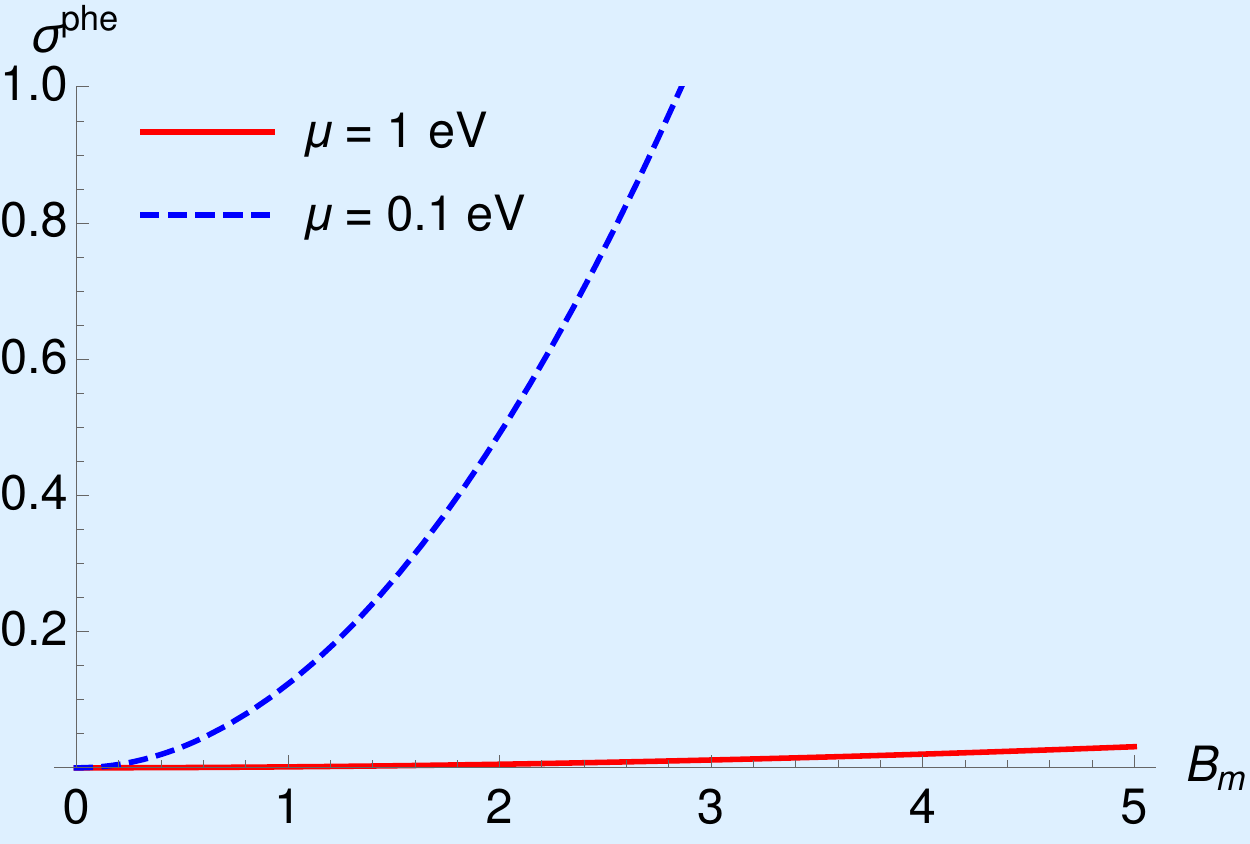}
\includegraphics[width=0.59\linewidth,height= 1.2 in]{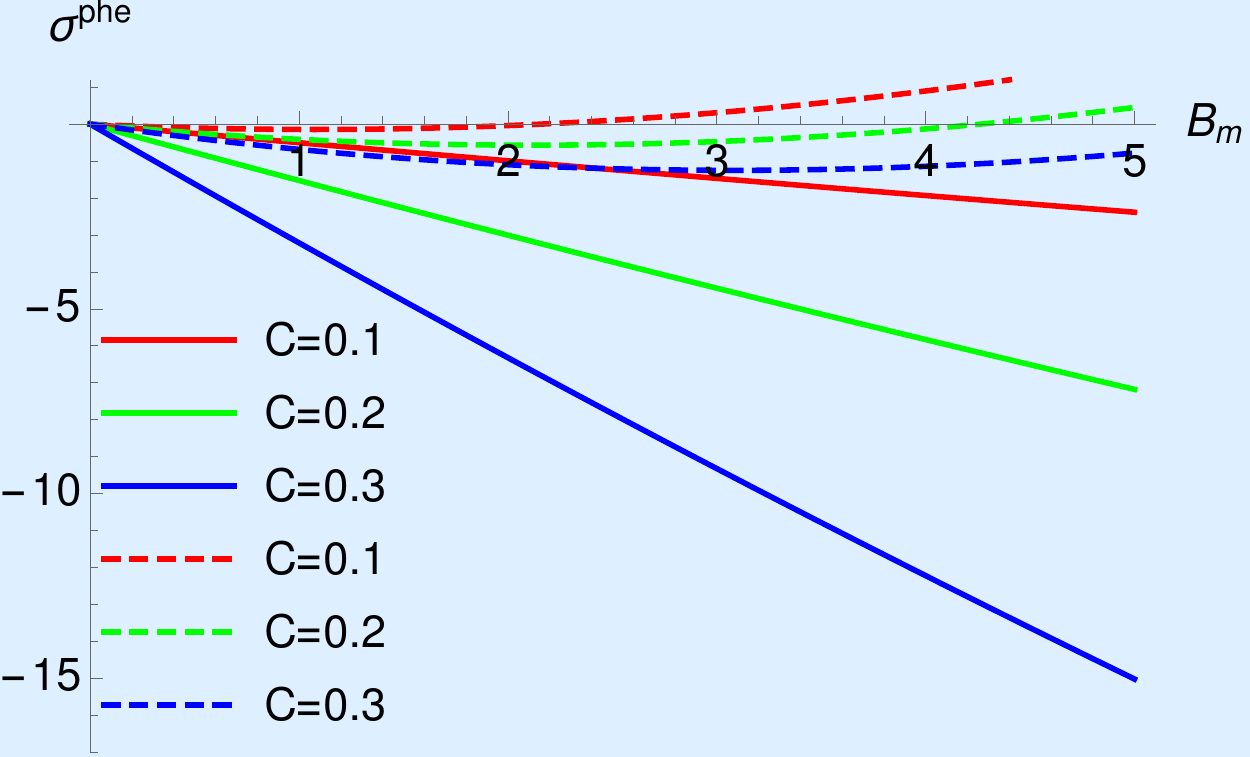}
\caption{(Color online) Variation of $\sigma^{phe}$ with $B_m$ at $\omega t=\pi/4$ for (left) $C=0$ and (right) $C\ne0$ (Dashed lines are for $\mu=0.1$ eV). All $\sigma_L$'s  are normalized by $10^4$ Siemens/m.} 
\label{xy-vs-B}
\end{figure}

For $C\ne0$, $\sigma^{phe}$ becomes linear in $B$ for $\theta_{eb}=0$ or $\pi/2$. For other angles, linear $B$ dependence is expected for small $B$ values if the strength of linear term is large compared to the quadratic term. Fig.\ref{xy-vs-B} shows how a change in $\mu$ from 1 eV to 0.1 eV can change the nature of $\sigma^{phe}$ field dependence from linear to quadratic. A temporal variation causes $\theta_{eb}$ to change and we witness a change in conductivity and its dependence on $B$. 
\begin{figure}
  \vskip .1 in
  \includegraphics[width=.48\linewidth,height=1.2 in]{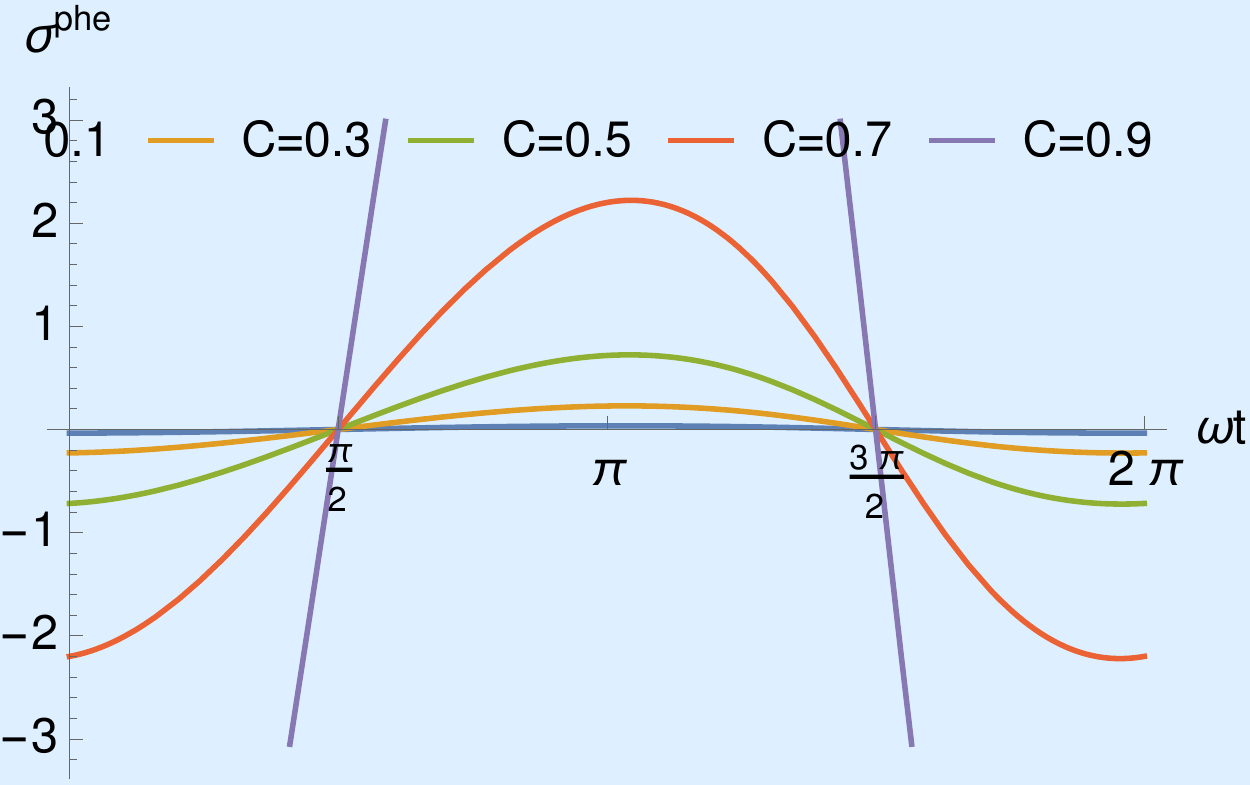}
  \includegraphics[width=.48\linewidth,height=1.2 in]{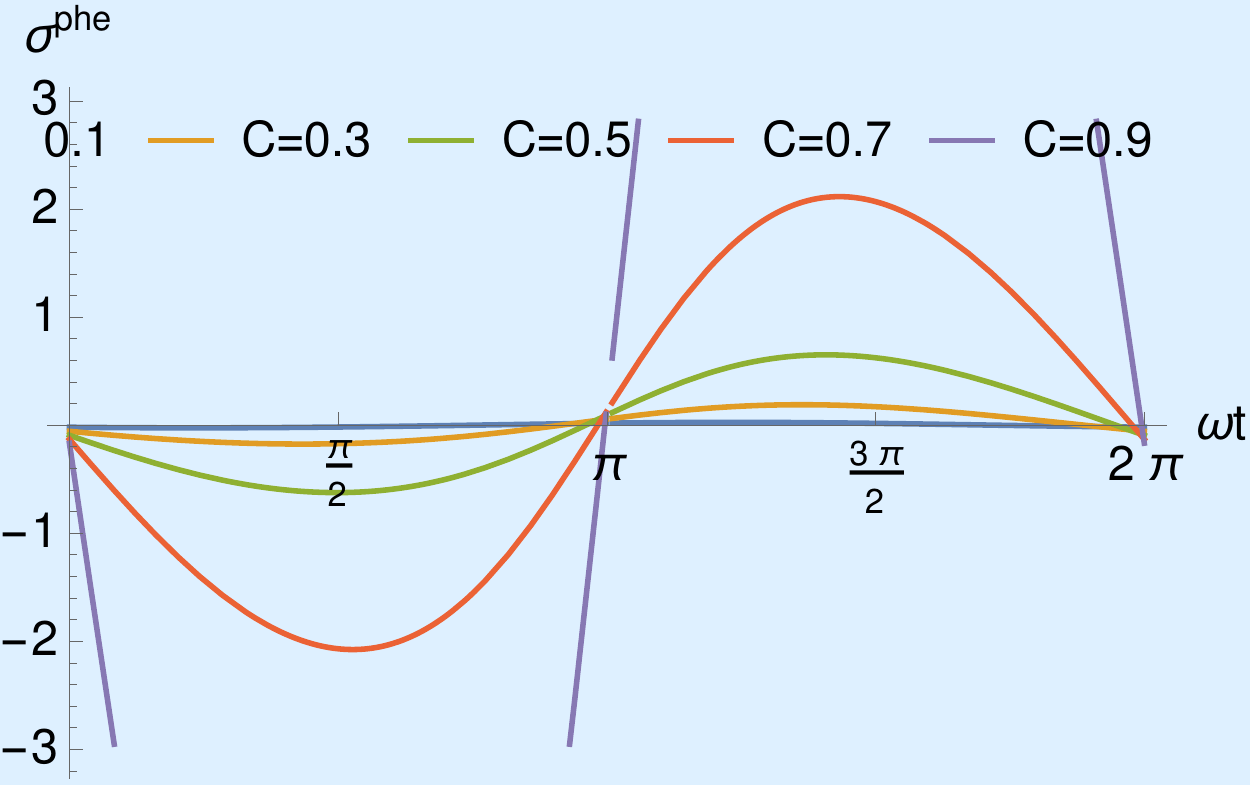}
  \includegraphics[width=.59\linewidth,height=1.2 in]{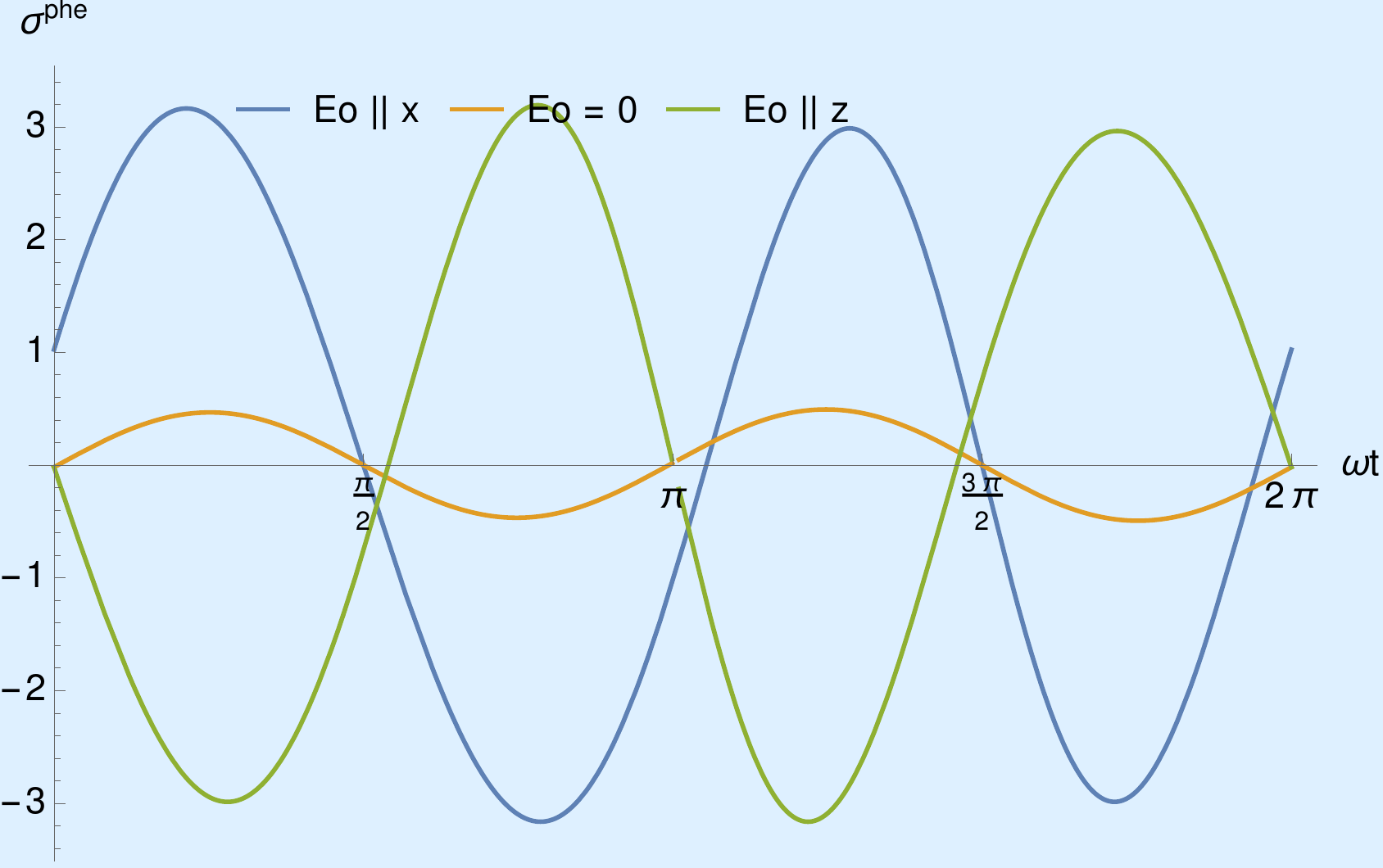}
\caption{(Color online) Variation of $\sigma^{phe}$ with $\omega t$ at B = 5 T for (bottom, with $\sigma^{phe}$ normalized by $10^2$ S/m) $C=0$ and (top, with $\sigma^{phe}$ normalized by $10^6$ S/m) $C\ne0$ [with $E_0$ along $x$ (top-left) and $z$ (top-right) respectively].} 
\label{xy-vs-t}
\end{figure}

Fig.\ref{xy-vs-t} shows the variation of $\sigma^{phe}$ with $\omega t$. Like LMC, the planar Hall effect also become stronger with tilting though the conductivity is not symmetric about $\omega t=0$, directly demonstrating the time reversal breaking. Interestingly, $\sigma^{phe}$ becomes zero at four (two) times within a cycle for $C=0$ ($C\ne0$). Those pair of points are same for all nonzero values of $C$ but changes with the strength of $B_m$. Moreover, if we consider $E_0$ to be along $\hat z$, the $\sigma^{phe}$-$~\omega t$ plot shows mostly a $\pi/2$ phase shift. Without $E_0$ however, the strength of $\sigma^{phe}$ diminishes.

\begin{figure}
\label{}
\end{figure}
\begin{figure}[t]
\includegraphics[width=.49\linewidth,height=1.4 in]{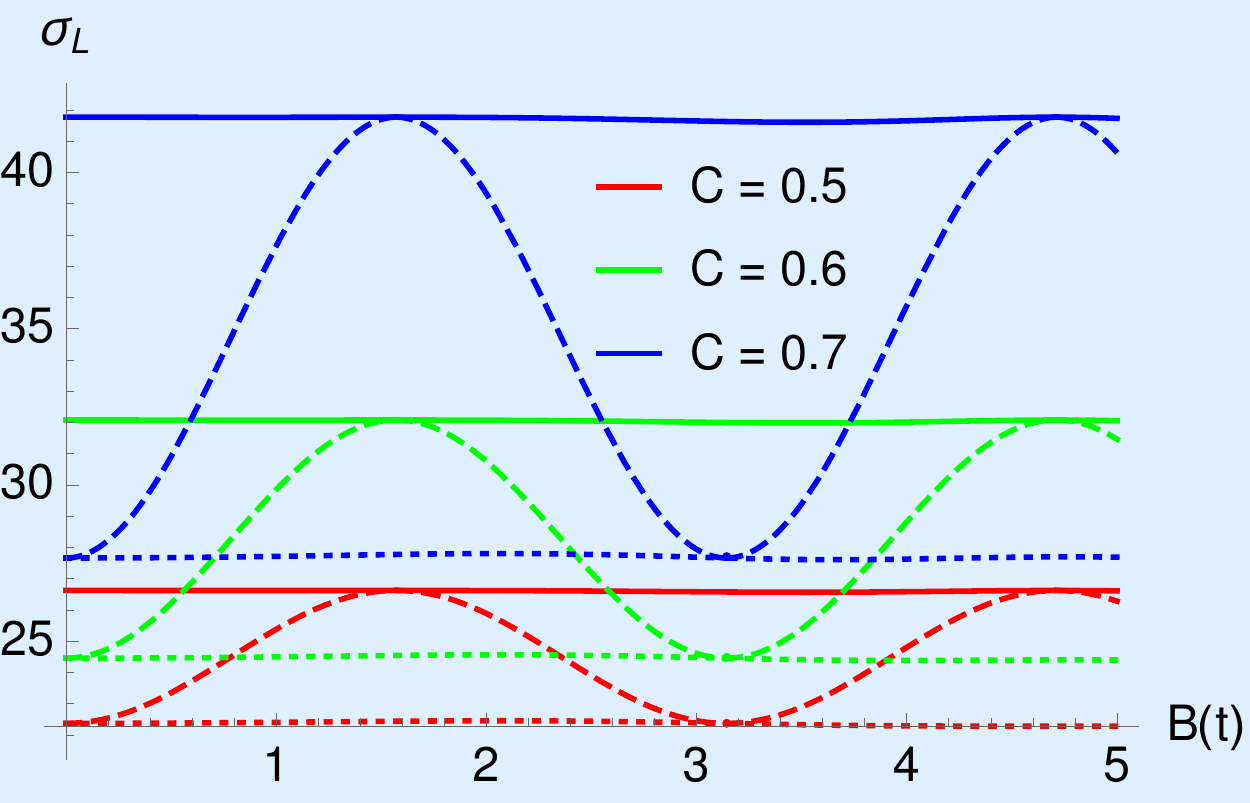}
\includegraphics[width=0.49\linewidth,height=1.4 in]{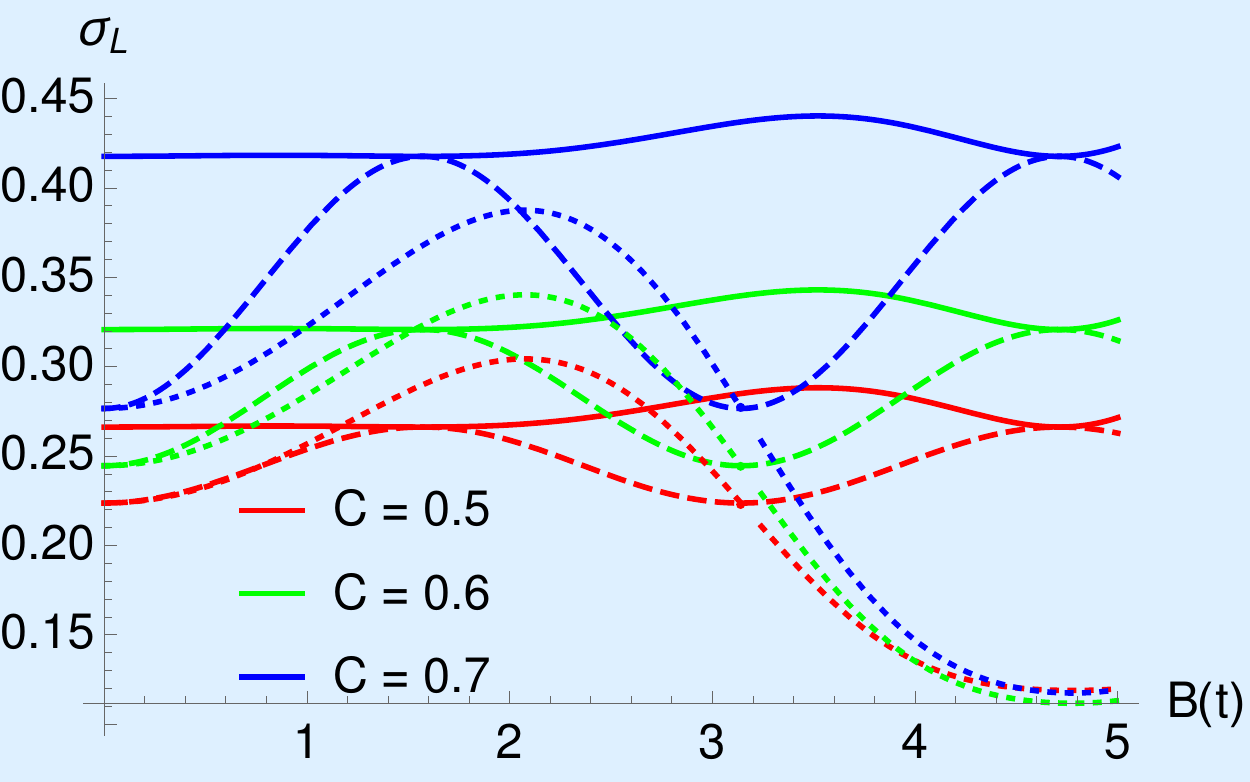}\\
\includegraphics[width=.49\linewidth,height=1.4 in]{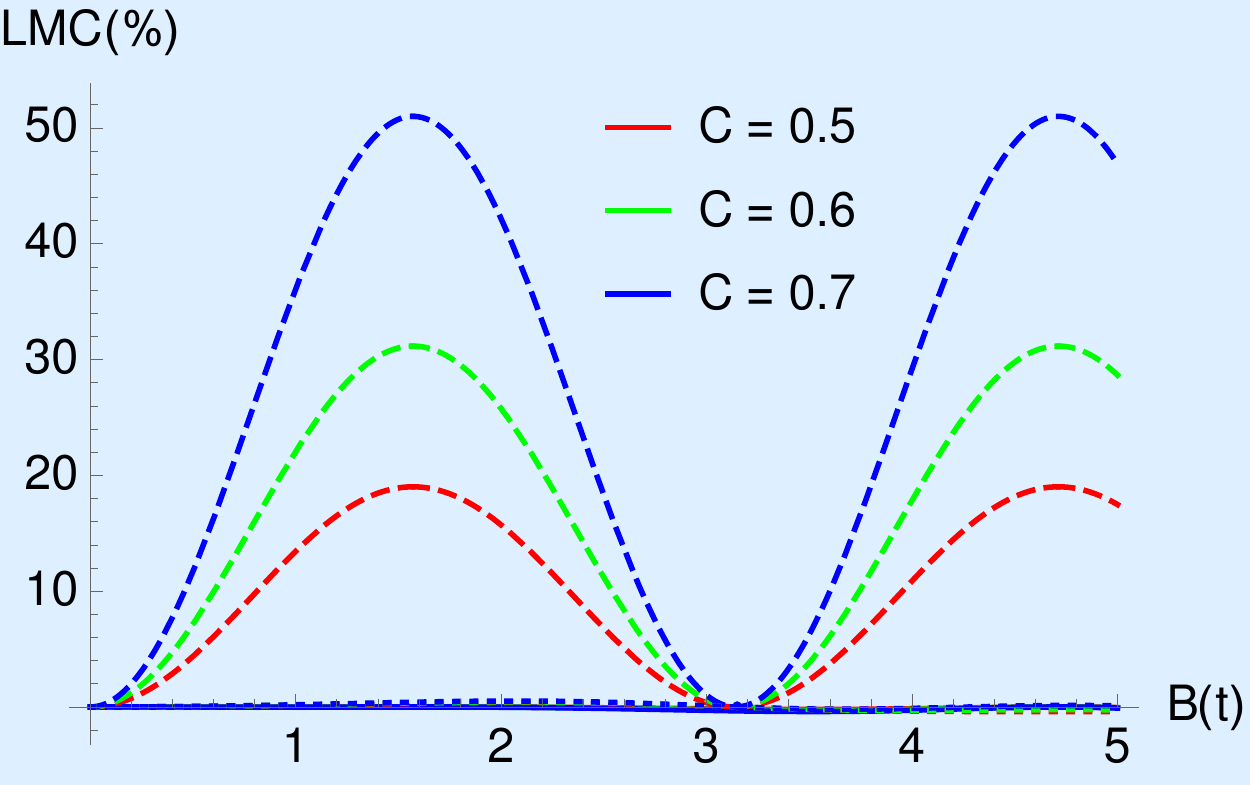}
\includegraphics[width=0.49\linewidth,height=1.4 in]{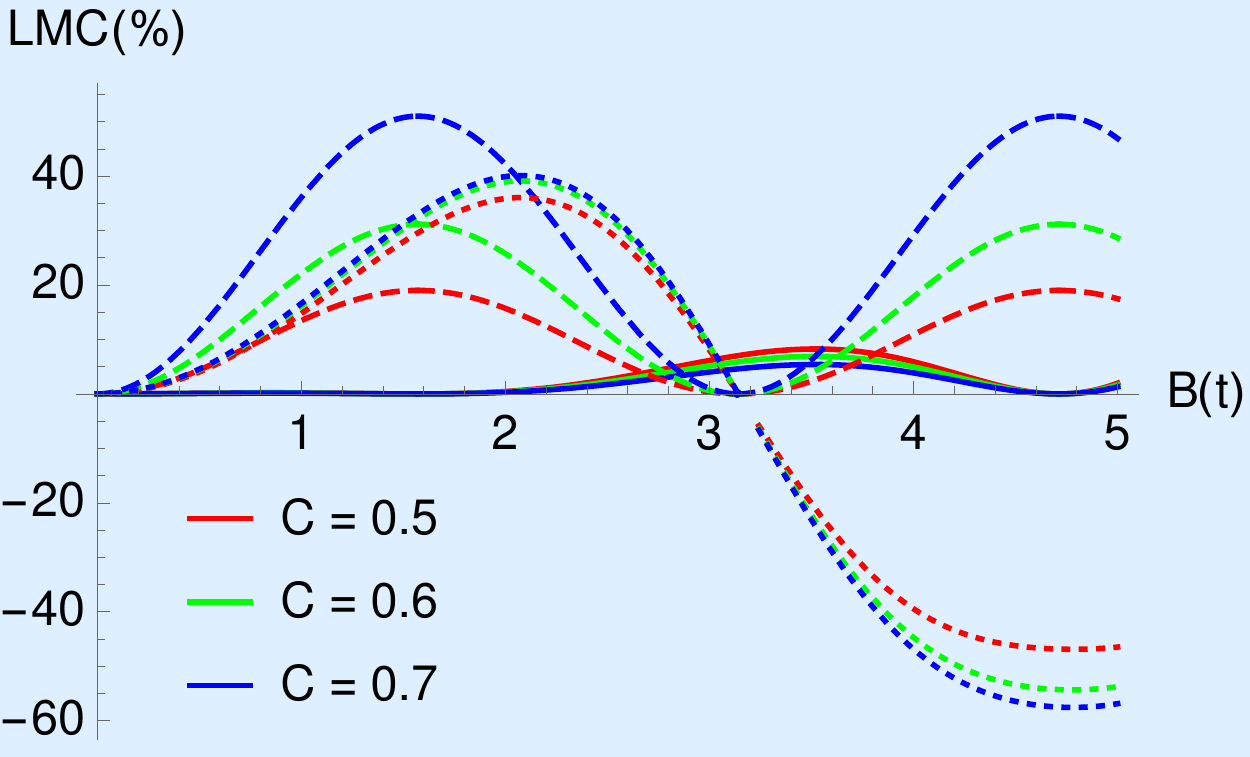}
\caption{(Color online) Variation of $\sigma_L$ with time dependent field $B(t)$ for $E_0||x$ (solid lines), $E_0=0$ (dashed lines) and $E_0||z$ (dotted lines) for different tilting parameters ($\sigma_L$'s normalized by $10^6$ S/m). We consider $\mu = 1$ eV (top-left) and $\mu = 0.1$ eV (top-right) respectively. {Corresponding LMCs are shown in bottom-left ($\mu = 1$ eV) and bottom-right ($\mu = 0.1$ eV) panels respectively.}} 
\label{sigma-Bt}
\end{figure}
\subsection{Time Varying irradiated field strengths}
In our chosen field set up, field strengths $E_m$ and $B_m$ are constants and so the field variations, that we have shown so far, can be compared experimentally by setting up different field amplitudes individually for obtaining different data points. Thus even though the results give good theoretical knowledge of how the conductivities should behave, it becomes difficult to observe them experimentally. {For example, in obtaining the results depicted in Fig.\ref{EozLMC2}, one has to set the fields to different values separately and wait for time $t=\pi/4\omega$ to measure $\sigma_L$.}
To avoid such cumbersome process, we can consider simple field amplitude variations such as $B_m=\alpha t$ (and thus $E_m=c\alpha t$). That way one can observe the variations of magnetoconductivities with time or the time dependent field $B(t)$ and this can be easily designed in labs for comparison purpose. In Fig.\ref{sigma-Bt}, we have shown such variations for $\mu=1$ and $0.1$ eV where we took $\alpha=\omega$ numerically. Notice that in absence of magnetic field at $t=0$, there is no chiral anomaly. {In this linear response theory, $\sigma_L$ doesn't depend on the strength of $E_0$ but depends on its direction. We see that $\sigma_L$ increases if the orientation of the static field changes from nodal direction $\hat z$ (with $\theta_L=\pi/2$) to the normal direction $\hat x$ (with $\theta_L=0$).  Then for $t\ne0$, the system witnesses fluctuating behavior in $\sigma_L$. In Fig.\ref{sigma-Bt} bottom-panels, we show how LMC (which is $\frac{\sigma_L(B)-\sigma_L(B=0)}{\sigma_L(B=0)}\times100\%$) can become positive and negative (for $E_0\ne0$) depending on the strength of $B(t)$.} Particularly for small energies, for example with $\mu=0.1$ eV, negative LMC can be clearly observed only when the static field acts along $\hat z$ (or, has nonzero components along it). Thus we can easily tune the system from the nontrivial positive LMC regime to trivial negative LMC regime and vice versa.

\section{Summary}

In this paper, we have theoretically studied the effect of an oscillating field on the magneto-transport in a WSM system {and analyzed the} features in electrical conductivities. {Our field set up causes continuous change in $\theta_{eb}$ with time that leads to nontrivial changes in the conductivities.} Unlike in metals, LMC in WSM usually take positive values. But here we find negative LMCs as well for few finite tilting parameter regime within the type-I tilting limit. Longitudinal conductivity variation with field shows linear or quadratic dependences depending on $C$ or $B_m$ values as well as on Fermi level $\mu$. Similar {results} are reported for planar Hall conductivities as well.
In this paper we also propose a temporal variations in ac field amplitudes where the field strengths are gradually turned on with time. This can be easily designed in experimental labs and its nontrivial outcomes of magnetoconductivities can be tested and then utilized to serve many purposes. {In Fig.\ref{sigma-Bt}, one witnesses the fluctuating behavior of $\sigma_L$ when the time dependent field strength is turned on gradually.} Though there are many studies of magneto-transport in Weyl systems under periodic driving\cite{ipsita1,ipsita2,banasri}, {the present study uses a combination of dc and ac fields which is first of its kind}, as per the best of the authors' knowledge. Based on this preliminary results and discussions, one can always build up further understanding incorporating additional complexities like using fully anisotropic BTE for considering anisotropy due to coupling between ${\bf B}$ and ${\bf \Omega}$\cite{park,johansson}, considering the effect of intra-node scattering\cite{girish20} or incorporating the orbital magnetic moments\cite{girish20,gao} etc. {Dynamics of our periodically driven problem can be analyzed using adiabatic-impulse approximation or Floquet formalisms for small and high frequency quenches respectively\cite{kar,ipsita1,banasri}.} One can also probe the thermal and Nernst conductivities under the same field set up\cite{banasri,tanay1} or investigate the response from multi-Weyl systems\cite{ipsita1,banasri,tanay1,tanay2}.
We should mention here that the tendency of $\sigma_L$ to become singular as $C\rightarrow1$ is an artifact of using the continuum model and {we have plans to later use a more realistic lattice model possibly incorporating beyond nearest-neighbor electron hoppings and reinvestigate this problem.} This will also enable us the study the response from a type-II WSM system\cite{prl119}.

\section*{Acknowledgements}
SK thanks G. Sharma, {B. Basu, S. De}, C. S. Yadav, D. Sinha and {A. Menon} for fruitful discussions.
This work is financially supported by DST-SERB, Government of India under grant no. SRG/2019/002143.



\end{document}